\documentclass[twocolumn]{aastex63}
\usepackage{amsmath}
\usepackage{xcolor}
\usepackage{lineno}

\newcommand\vsh{v_\mathrm{sh}}
\newcommand\va{v_\mathrm{A}}
\newcommand\ncr{N_\mathrm{cr}}
\newcommand\gcr{\Gamma_\mathrm{cr}}
\newcommand\np{N_\mathrm{p}}
\newcommand\ucr{U_\mathrm{cr}}
\newcommand\ubu{U_\mathrm{bulk}}
\newcommand\omp{\Omega_\mathrm{p}}
\newcommand\opp{\omega_\mathrm{p,p}}
\newcommand\ma{M_\mathrm{A}}

\newcommand{\mpo}{\textcolor{black}}
\newcommand{\mpon}{\textcolor{black}}
\newcommand{\mponn}{\textcolor{black}}
\newcommand{\mpf}{\textcolor{black}}

\submitjournal{ApJ}

\shorttitle{Cosmic-ray spectral steepening}
\shortauthors{Pohl}

\begin{document}

\title{Time-dependent treatment of cosmic-ray spectral steepening due to turbulence driving}

\email{marpohl@uni-potsdam.de}

\author[0000-0001-7861-1707]{Martin Pohl}
\affiliation{University of Potsdam, Institute of Physics and Astronomy, D-14476 Potsdam, Germany}
\affiliation{DESY, D-15738 Zeuthen, Germany}

\begin{abstract}
Cosmic-ray acceleration at non-relativistic shocks relies on scattering by turbulence that the cosmic rays drive upstream of the shock. We explore the rate of energy transfer from cosmic rays to non-resonant Bell modes and the spectral softening it implies. Accounting for the finite time available for turbulence driving \mponn{at supernova-remnant shocks} yields a smaller spectral impact than found earlier with steady-state considerations. Generally, for diffusion scaling with the Bohm rate by a factor $\eta$, the change in spectral index is at most $\eta$ divided by the Alfv\'enic Mach number of the thermal sub-shock. For \mponn{$\ma\lesssim 50$} it is well below this limit. Only for very fast shocks and very efficient cosmic-ray acceleration the change in spectral index may reach $0.1$. For standard SNR parameters it is negligible. \mponn{Independent confirmation is derived by considering the synchrotron energy losses of electrons: if intense nonthermal multi-keV emission is produced, the energy loss, and hence the spectral steepening, is very small for hadronic cosmic rays that produce TeV-band gamma-ray emission.}
\end{abstract}

\keywords{Gamma-ray astronomy --- cosmic rays --- supernova remnants --- spectral index}

\section{Introduction} \label{sec:intro}
\citet{2019MNRAS.488.2466B} studied cosmic-ray acceleration at nonrelativistic shocks using a tensor expansion of the Vlasov equation, and they found a steepening of the cosmic-ray spectrum \mponn{at supernova remnants (SNR)} arising from energy transfer from the cosmic rays to turbulence in the precursor of the shocks. Assuming the turbulence in question is nonresonant, the so-called Bell mode \citep{2004MNRAS.353..550B}, and taking an estimate for its \mponn{magnetic-field} energy density at the saturation level, 
\begin{equation}
U_{\delta B} \approx \frac{\vsh}{2c}\,U_\mathrm{cr},
\label{eq1}
\end{equation}
where $\vsh$ denotes the shock speed and $U_\mathrm{cr}$ the energy density in cosmic rays immediately upstream of the shock, they find a spectral steepening by 
\begin{equation}
\Delta s\vert_\mathrm{Bell}\approx \frac{4}{\epsilon}\frac{U_{\delta B}}{U_\mathrm{cr}}
\approx \frac{2}{\epsilon}\frac{\vsh}{c} .
\label{eq:bell1}
\end{equation}
\mponn{For young SNR this spectral change should be quite sizable.}
Here $\epsilon$ is the fraction of turbulent energy density that is in the form of magnetic-field \mponn{fluctuations, $U_{\delta B}=\epsilon\,U_\mathrm{wave}$}. \mponn{For an incompressive linear wave with phase speed $v_\phi < \va$, like Bell's mode, we have $\epsilon\gtrsim 0.5$.} Clearly, in the nonlinear phase quite a bit of energy is transferred to velocity fluctuations \citep{2009ApJ...706...38S} and local displacements of electrons and ions \citep{2017MNRAS.469.4985K}, suggesting a moderate variation in $\epsilon$. The modification of the bulk flow speeds \citep{2009MNRAS.397.1402L,2017MNRAS.469.4985K,2019ApJ...873...57W} is a nonlinear transfer of wave energy to the plasma, as are secondary instabilities that thrive on Bell modes \citep{2011MNRAS.410...39B}. \mpf{These processes} do not change $\epsilon$, and so $\epsilon=0.5$ should be approximately valid for Bell modes in the nonlinear stage.

The result conforms with intuition. The streaming of cosmic rays is the energy supply of the turbulence. In the steady state the production rates of turbulence and cosmic rays in the upstream region and at the shock must balance the loss incurred by advection to the \mpf{far-downstream} region. The advection speed is the same for the two. The energy density of the turbulence is increased by compression at the shock, and so the fraction of the cosmic-ray production power that is funneled to turbulence is a few times the ratio of their upstream energy density, $U_{\delta B}/\epsilon U_\mathrm{CR}$. The cosmic-ray spectrum reflects the balance between the energy gain and the retention probability for each acceleration cycle \citep{1978MNRAS.182..147B}. Energy transfer to turbulence reduces the energy gain of cosmic rays, hence the spectral steepening. Equation~\ref{eq:bell1} then corresponds to the highest possible effect, given the constraint $k r_\mathrm{L}\gg 1$ for Bell's instability.

\mpo{What of the uncertainties in the estimate? Does turbulence driving perhaps impose termination of acceleration as opposed to a continuous spectral steepening? Which fraction of the observed post-shock magnetic energy density is actually carried by Bell's mode, and \mpf{what part is} provided by other processes operating at the shock \citep[e.g.][]{2007ApJ...663L..41G}? Do the waves at the shock fully sample the cosmic-ray energy loss or have the waves lost energy already in the upstream region? And} is the analytically estimated saturation level really relevant? Is there enough time for Bell's mode to grow to very high amplitude in the precursor of a shock, before the plasma arrives at the shock and the driving of waves stops? For simplicity we consider cosmic rays with Lorentz factor $\gcr$. The peak growth rate of the mode scales with the proton gyrofrequency, $\omp$, as
\begin{equation}
\gamma_\mathrm{max}\simeq \omp\,\frac{\vsh\,\ncr}{2\,\va\,\np}=\opp\,\frac{\vsh\,\ncr}{2\,c\,\np} ,
\label{eq2}
\end{equation}
where $\va$ \mpf{is} the Alfv\'en speed, \mponn{$\opp$ stands for the proton plasma frequency, and $\ncr$ and $\np$ \mpf{denote} the number density of cosmic rays and of ambient protons, respectively}. 
 
The cosmic-ray precursor has an extent that is determined by the ratio of the diffusion coefficient, here written as multiple of the Bohm diffusion coefficient, $\kappa=\eta r_\mathrm{L} c/3$, and the shock speed, $\vsh$. Strong driving of turbulence or cosmic-ray feedback can moderately modify, and in situations with efficient particle acceleration reduce, the extent. Magnetic-field amplification reduces the Larmor radius, and \mponn{\citet[e.g.][]{2013MNRAS.430.2873R} find $\kappa$ to be well described by} either $\eta$ \mpf{being} a few for $r_\mathrm{L}$ as measured immediately upstream of the shock or $\eta\lesssim 1$ with the far-upstream value of $r_\mathrm{L}$. \mponn{In this paper we shall \mponn{use the former description and denote} with $\delta B$ the amplitude of turbulent magnetic field very close to the shock, unless it is explicitely written as $\delta B(r)$. Following the simulation results of \citet{2013MNRAS.430.2873R} we shall use $\eta=4$ as \mpf{a} fiducial value and consider deviations from that in the discussion section.}
The shock-capture time needed for the shock to cross a region as wide as the shock precursor is
\begin{equation}
\tau_\mathrm{sc} \approx \frac{\kappa}{\vsh^2}=
\frac{\eta\,\gcr\,c^2}{3\,\omp\,\vsh^2} .
\end{equation}
The number of exponential growth cycles available for Bell's mode then is \citep{2008ApJ...684.1174N}
\begin{equation}
N_\mathrm{exp} =\gamma_\mathrm{max} \tau_\mathrm{sc}\approx \frac{\eta\gcr c^2\ncr}{6\,\va\vsh\np}=\frac{\eta\ma}{12}\frac{\ucr}{\ubu},
\label{nexp}
\end{equation}
where $\ma$ is the Alfv\'enic Mach number of the shock, and $U$ denotes the energy density in cosmic rays and in the bulk plasma flow ($=1/2\,\rho \vsh^2$). Efficient cosmic-ray acceleration may funnel $10\%$ of the bulk-flow energy \mpf{into} cosmic rays. Given our expectation $\eta\approx 4$, it becomes clear that one needs Alfv\'enic Mach numbers of $\ma \approx 300$ or more to allow 10 exponential growth cycles of the mode. \mponn{Note that Bell's mode is perceived to grow out of the fluctuation spectrum in the interstellar medium or in the wind bubble of a core-collapse supernova, and \mpf{it is supposed} to reach an amplitude much larger than that of the large-scale field, $\delta B \gg B_0$, which realistically takes an amplitude enhancement by much more than a factor thousand (or more than seven exponential cycles). We note that even the notoriously noisy particle-in-cell simulations require ten growth cycles, if not more, to reach significant magnetic-field amplification \citep{2009ApJ...694..626R,2009ApJ...706...38S,2010ApJ...711L.127G,2017MNRAS.469.4985K}}. For a low upstream gas density, $\np\lesssim 0.5\ \mathrm{cm}^{-3}$, and standard values of the \mponn{interstellar magnetic field, $B=7\ \mathrm{\mu G}$ \citep{2012ApJ...761L..11J}}, this is difficult to establish, \mpf{on account of} $\va \gtrsim 20\ \mathrm{km\,s^{-1}}$. \mponn{In the environment of a core-collapse supernova the magnetic field is less well known \citep{2015SSRv..191..339R}, but should eventually scale inversely with the radius, $B\propto 1/r$ \citep{1982ApJ...253..188V}, implying a spatially constant Alfv\'en speed in the free-wind zone. Whatever the value of the Alfv\'en speed far upstream of the shock, significant magnetic-field amplification in the cosmic-ray precursor will reduce it.} For high-density environments damping may strongly limit the growth of the mode \citep{2007A&A...475..435R}. \mpon{Reaching \mponn{a sufficient number of growth times for magnetic-field amplification} is certainly possible under specific conditions, but not something one should expect for any SNR. If the conditions are met, then for only a short period in the evolution of the remnant, because the outer shock is likely too slow already at the end of the free-expansion phase.}

\citet{2020A&A...634A..59B} demonstrated for Alfv\'enic turbulence that slow build-up of turbulence in the upstream region will rapidly reduce the maximum energy, to which an SNR can accelerate. In any case, if turbulence driving would significantly soften the cosmic-ray spectrum, $\ucr$ would fall off with increasing energy. At very high energy only very few growth cycles would be available and/or $\eta$ would be very large, both of which would stymie further acceleration.

In this paper we present estimates of the energy loss of cosmic rays to turbulence that account for the limited time available for the development of the nonresonant streaming instability. \mponn{In our discussion we shall use parameter values that may represent the forward shocks of young SNRs, but in principle the analysis applies to any fast non-relativistic shock at which diffusive shock acceleration operates.}

\section{Time-dependent modeling of turbulence build-up} 
We shall explore two independent ways to estimate the energy transfer from cosmic rays to turbulence. One will be a local consideration of turbulence driving in the precursor, the other an integral assessment of the energy transfer in the entire precursor. For simplicity we shall assume that all ions in the cosmic rays and the background plasma are protons, i.e. hydrogen nuclei. The non-resonant and broadband character of Bell's mode permits considering cosmic rays of a specific energy as \mpf{a} proxy for particles in a wide energy band.

\subsection{Local estimate} \label{sec2.1}
The growth rate of wave energy density is proportional to the \mpf{energy density} in the unstable wave band,
\begin{equation}
\dot U \simeq \int dk\ \gamma(k)\,\frac{B_k^2}{4\pi\epsilon}\lesssim \gamma_\mathrm{max}\,\frac{(\delta B)^2}{4\pi\epsilon}=
\gamma_\mathrm{max}\,\frac{2\,U_{\delta B}}{\epsilon},
\label{eq3}
\end{equation}
where again $\epsilon$ is the fraction of the wave energy that is carried by magnetic field, $\delta B$ is the magnetic wave amplitude, \mponn{and $U_{\delta B}$ the magnetic-field energy density of the waves.} Equation~\ref{eq3} indicates that most of the energy transfer arises when the wave amplitude is high, meaning near or at saturation. \mpon{The expression is written as \mpf{an} upper limit that is conservative for two reasons. First, it applies the peak growth rate to the entire wave spectrum, including spectral bands in which only cosmic rays of vastly different energy can drive Bell's mode. Second, in the highly nonlinear phase the growth rate, $\gamma_\mathrm{max}$, might be lower than it is for moderate $\delta B$. Both effects can only strengthen the validity of our upper limit for the growth rate of wave energy density.}
Inserting Eq.~\ref{eq2} gives\mpo{
\begin{equation}
\dot U \lesssim \frac{\opp}{\epsilon} \frac{\ncr}{\np}\frac{\vsh}{c} U_{\delta B}.
\label{eq4}
\end{equation}
Here $\opp$ is the ion (proton) plasma frequency. To be noted from eq.~\ref{eq4} is that the energy\mpf{-density} transfer rate is independent of the amplitude of the large-scale magnetic field, $B_0$. It is evident that most of the energy transfer occurs deep in the cosmic-ray precursor and close to the shock, where both $\ncr$ and $U_{\delta B}$ reach their upstream peak values.}

\mpf{The} energy gain by the waves imposes energy loss \mpf{on} the cosmic rays. We can estimate the energy-loss time per cosmic-ray particle as\mpo{
\begin{equation}
\tau_\mathrm{loss} \simeq \frac{\ucr}{\dot U} \gtrsim \frac{2\epsilon\gcr}{\opp}\frac{\ubu}{U_{\delta B}}\frac{c^3}{\vsh^3},
\label{eq5}
\end{equation}
which evidently is independent of the number density of cosmic rays.}
The only property of the cosmic rays that enters Eq.~\ref{eq5} is their Lorentz factor. 

\mpo{We shall now compare the energy-loss time (Eq.~\ref{eq5}) with the acceleration time assuming diffusive shock acceleration. As the nonresonant streaming instability operates only upstream of the shock, our comparison can ignore the part of the acceleration time that is spent downstream. Particles \mpf{spend} approximately half of their time in the upstream region, and so the effective energy-loss time is twice that given in eq.~\ref{eq5},
\begin{equation}
\tau_\mathrm{loss, eff} \simeq 2\tau_\mathrm{loss} \gtrsim \frac{4\epsilon\gcr}{\opp}\frac{\ubu}{U_{\delta B}}\frac{c^3}{\vsh^3}.
\label{eq5a}
\end{equation}}
Again expressing the diffusion coefficient of relativistic cosmic rays in Bohm units, $\kappa=\eta c\,r_\mathrm{L}/3$, we rewrite equation (32) of \citet{1991MNRAS.251..340D} as\mpo{
\begin{equation}
\tau_\mathrm{acc} = \frac{8\,\kappa}{\vsh^2} = \frac{8\eta \gcr}{3\omp }\frac{c^2}{\vsh^2}.
\label{eq6}
\end{equation}}
Following \citet{1978MNRAS.182..147B}, the integral cosmic-ray spectrum reflects the balance between acceleration and escape by advection to the far downstream region, hence for the spectral index $s-1=\tau_\mathrm{acc}/\tau_\mathrm{esc}$. Energy losses by driving turbulence increase the effective acceleration time scale, leading to a softened spectrum
\begin{equation}
\frac{dN(>E)}{dE}\approx \frac{\Delta N(>E)}{\Delta E}\simeq \frac{1-s}{1-\frac{\tau_\mathrm{acc}}{\tau_\mathrm{loss, eff}}}\,\frac{N(>E)}{E}\ .
\label{eq6a}
\end{equation}
For test particles at a strong shock in \mpf{a} hydrogen plasma one expects $s=2$. The ratio of timescales in the denominator of eq.~\ref{eq6a} must be less than unity, otherwise acceleration is impossible. For a small ratio of acceleration time and loss time the resulting change of spectral index is\mpo{
\begin{align}
\Delta s&= \frac{s-1}{\frac{\tau_\mathrm{loss, eff}}{\tau_\mathrm{acc}}-1}\simeq
(s-1) \frac{\tau_\mathrm{acc}}{\tau_\mathrm{loss,eff}}\quad \Rightarrow  \nonumber \\
\Delta s&\lesssim \frac{2\,(s-1)\,\eta}{3\,\epsilon}\frac{U_{\delta B}}{\ubu} \frac{\vsh}{c}\frac{\opp}{\omp}.
\label{eq7}
\end{align}
Here the proton gyrofrequency, $\omp$, derives from the scaling of the acceleration time with the Bohm diffusion coefficient. \citet{2013MNRAS.430.2873R} noted that for efficient magnetic-field amplification \mponn{the diffusion coefficient is well described by} the amplified field strength and $\eta\approx 4$. One can rewrite expression \ref{eq7} \mponn{using the Alfv\'enic Mach number, $\ma=\vsh\opp/c\omp $,} that is calculated with the amplitude of the amplified field without regard of direction,
\begin{equation}
\Delta s\lesssim \frac{2\,(s-1)\,\eta\,\ma}{3\,\epsilon}\frac{U_{\delta B}}{\ubu} .
\label{eq7a}
\end{equation}}
\subsection{Global assessment}\label{sec2.2}
We shall now conduct a global assessment of the cosmic-ray energy loss incurred in the entire 
upstream region. The spatial profile of cosmic-ray density in the so-called precursor is dominated by the homogeneous solution to the spatial part of the transport equation,
\begin{equation}
\ncr \simeq N_\mathrm{cr,sh} \exp\left(-\int_{r_\mathrm{sh}}^r\ \frac{dr^\prime\ v(r')}{\kappa(r')}\right),
\label{eq8}
\end{equation}
\mpo{where $v(r)$ is the upstream flow speed measured in the shock rest frame.} \mpon{This steady-state profile represents a balance between the advective flux toward the shock and \mpf{the} diffusive flux away from it. The diffusive flux, $-\kappa\,\partial\ncr/\partial r$, also determines the cosmic-ray current and hence the driving rate of Bell's mode. If the cosmic-ray gradient were weaker than its steady-state value, \mpf{cf.} Eq.~\ref{eq8}, then advection would dominate the spatial transport, and the density profile would evolve toward the steady state; likewise for a steeper density profile. Equation~\ref{eq8} hence represents a stable equilibrium and can be used in our subsequent calculations.}

\mpon{The cosmic-ray density profile also determines the acceleration timescale, because that depends on the average separation of upstream cosmic rays from the shock, $\langle r-r_\mathrm{sh}\rangle $. In each acceleration cycle, cosmic rays enjoy a relative energy gain on the order of $\vsh/c$ and need a few times $\langle r-r_\mathrm{sh}\rangle/c$ for it. Hence the acceleration time is a few times $\langle r-r_\mathrm{sh}\rangle/\vsh$. If $v$ and $\kappa$ were constant, then 
\begin{equation}
\langle r-r_\mathrm{sh}\rangle= \frac{\kappa}{v}=\frac{\kappa}{\vsh}\, ,
\label{eq8h}
\end{equation}
and we recover the formula $\tau_\mathrm{acc}\propto \kappa/\vsh^2$ \citep{1991MNRAS.251..340D}. If $v/\kappa$ would significantly decrease at $r-r_\mathrm{sh}\gg \langle r-r_\mathrm{sh}\rangle$, then we would arrive at approximately the same conclusion, except that there might be some (weak) escape toward the far upstream. If $v/\kappa$ would significantly decrease already close to the shock, then $\langle r-r_\mathrm{sh}\rangle$ would be very large and may in fact become unbound, in which case many of the freshly accelerated cosmic rays escape to the far upstream, and the cosmic-ray spectrum would be very steep \citep{2020A&A...634A..59B}. The rapid increase of volume with increasing $r$ would contribute to the cut-off in the spectrum of confined particles \citep{2010A&A...513A..17O}.}

\mpon{In the steady state the cosmic-ray current would be unaffected and could still drive the non-resonant mode \citep{2009ApJ...694..951R}, thus reducing $\kappa(r)$ and leading to a seemingly universal profile of the cosmic-ray precursor \citep{2013MNRAS.431..415B}. In essence, a significant increase of $\kappa$ with distance from the shock would give ample time for turbulence driving, but would make particle acceleration slow and inefficient, and hence impose a cut-off in the spectrum. If that were the case, young SNRs could not have produced cosmic rays of very high energy. The TeV-band detection of many SNRs suggests otherwise, and so we will proceed with the assumption that the variation in $\kappa(r)$ does not preclude a rapid decline of the cosmic-ray density in the precursor. \mpon{In other words, we consider cosmic-ray energies well below the maximum the shock can presently provide.} Studies of cosmic-ray-modified shocks \citep{2006MNRAS.371.1251A} suggest that this approximation is good.}

We shall now explicitly consider variations in the flow speed and hence write eq.~\ref{eq2} with $v(r)$ instead of $\vsh$. Using eq.~\ref{eq3} we integrate the energy transfer rate over the entire precursor, which for a plane-parallel shock means integration over $r$,\mpo{
\begin{equation}
\dot E_\mathrm{tot} \lesssim \frac{N_\mathrm{cr,sh}}{8\pi\,\epsilon \, c} \int_{r_\mathrm{sh}}^\infty dr\ \frac{\omega_\mathrm{pp}\, v(r)\,(\delta B(r))^2}{\np\,
\exp\left(\int_{r_\mathrm{sh}}^r dr'\ \frac{v(r')}{\kappa(r')}\right)} .
\label{eq8a}
\end{equation}
Continuity mandates that $\omega_\mathrm{pp}/\np \propto \sqrt{v(r)}$. We can scale the density and the flow speed to their values at the thermal sub-shock, $N_\mathrm{p,sh}$ and $\vsh$, and pull these quantities out of the integral, 
\begin{equation}
\dot E_\mathrm{tot} \lesssim \frac{\omega_\mathrm{pp, sh}\, N_\mathrm{cr,sh}}{8\pi\,\epsilon \, c\,N_\mathrm{p,sh}\sqrt{\vsh}} \int_{r_\mathrm{sh}}^\infty dr\ \frac{v(r)^{3/2}\,(\delta B(r))^2}{
\exp\left(\int_{r_\mathrm{sh}}^r dr'\ \frac{v(r')}{\kappa(r')}\right)} .
\label{eq8aa}
\end{equation}
Note that $v(r)$ is measured in the sub-shock rest frame, and so non-linear cosmic-ray feedback would lead to a positive gradient in $v(r)$ and a negative gradient in $(\delta B(r))^2$ on account of compression, that partially compensate each other.}

The energy density in the magnetic turbulence likely increases toward the shock on account of turbulence driving, and so $(\delta B(r))^2$ falls off with increasing $r$. The cosmic-ray scattering rate is linked to the intensity of the turbulence, hence $\kappa(r)$ will rise with increasing $r$. We shall write the spatial profiles as
\begin{equation}
(\delta B(r))^2= (\delta B_\mathrm{sh})^2\, b(r)\qquad
\kappa(r)=\kappa_\mathrm{sh}\,k(r).
\label{eq8b}
\end{equation}
\mpon{Note that $\delta B(r)$ may include small-$k$ turbulence that is driven by cosmic rays of energies higher than that of the particles whose energy losses we calculate. The diffusion coefficient increases with energy, and so does the precursor length, implying that at any given location in the precursor a particle sees turbulence that has been driven by cosmic rays of higher energy further out in the precursor. Part of that turbulence has a scale commensurate with or larger than the Larmor radius of the particle in question, which provides additional scattering and possibly magnetic-field amplification, thus reducing $\eta$ in our description, but it does not necessarily impose additional energy loss on our particle.}

The variation of $b(r)$ and $k(r)$ may be huge, as we observe evidence of strong magnetic-field amplification. \mpon{We noted above that $\kappa(r)$ should not increase too quickly, otherwise the acceleration is inefficient and the cosmic-ray spectrum cuts off.} 

\mpo{We shall now \mponn{use the argument of the exponential in eq.~\ref{eq8aa} as new} variable of integration, 
\begin{equation}
x=\int_{r_\mathrm{sh}}^r dr' \,\frac{v(r')}{\kappa(r')} ,
\end{equation}
which \mponn{leads to}
\begin{align}
\dot E_\mathrm{tot} \lesssim\, &\frac{\omega_\mathrm{pp, sh}\, N_\mathrm{cr,sh}\,(\delta B_\mathrm{sh})^2\,\kappa_\mathrm{sh}}{8\pi\,\epsilon \, c\,N_\mathrm{p,sh}} \nonumber \\
 & \times \int_0^\infty dx\ \sqrt{\frac{v(x)}{\vsh}}\,b(x)\,k(x)\,
\exp(-x) ,
\label{eq8ab}
\end{align}
where $b(x)$ and $k(x)$ are defined in eq.~\ref{eq8b}. In the quasilinear limit $\kappa\propto 1/(\delta B)^2$ is approximately true, meaning $b(x)k(x)\approx\mathrm{const}$, whereas for Bohmian scaling one expects $b(x)k(x)\propto \sqrt{b(x)}$. In the general case, $b(x)k(x)$ is at most constant, but more likely a declining function, \mponn{because $b(x)$ is expected to decline,} and so the integral in eq.~\ref{eq8ab} yields a numerical factor close to unity.} \mpon{Explicitly considering the spatial variation of $\delta B$ and of the diffusion coefficient, $\kappa(r)$, thus leads to the same conclusion as the phenomenological analysis described in sec.~\ref{sec2.1}: a reasonable spatial \mponn{variation} of the relevant parameters does not invalidate the calculations presented in this section.} The total energy-loss rate of cosmic rays in the precursor then is
\begin{equation}
\dot E_\mathrm{tot} \lesssim  U_{\delta B} \frac{N_\mathrm{cr,sh}}{\np} \frac{\eta \gcr}{3\epsilon}\frac{c^2}{\va} ,
\label{eq9}
\end{equation}
\mpo{where we again used Bohm scaling for the diffusion coefficient, $\kappa=\eta r_\mathrm{L} c/3$.} 
The escape flux  to the far-downstream region,
\begin{equation}
\dot N_\mathrm{esc} =  - N_\mathrm{cr,sh} \frac{\vsh}{4},
\label{eq10}
\end{equation}
is in the steady state compensated by the energy gain on account of acceleration at the shock, and the level of balance determines the spectral index, s,
\begin{equation}
\dot E_\mathrm{acc} = \frac{m_p c^2 \gcr}{s-1} \vert\dot N_\mathrm{esc} \vert\ .
\label{eq10a}
\end{equation}
The ratio of the two rates in eqs.~\ref{eq9} and \ref{eq10a} is a measure of the spectral steepening,  
\begin{equation}
\Delta s=(s-1) \frac{1}{\frac{\dot E_\mathrm{acc}}{\dot E_\mathrm{tot}}-1}
\lesssim\frac{2\,(s-1)^2\,\eta\,\ma}{3\,\epsilon}\frac{U_{\delta B}}{\ubu}.
\label{eq11}
\end{equation}
As for eq.~\ref{eq7}, we expanded this formula for weak losses to derive the last expression. Apart from a factor $s-1$, we find the same level of spectral steepening as in eq.~\ref{eq7}. There we had used a formula for the acceleration time scale that was strictly derived for test particles at a strong shock, i.e. $s=2$, for which there is no difference between the two estimates for the spectral steepening. \mpo{We remind the reader that we integrated out spatial variations in the cosmic-ray precursor and that all variables are supposed to be measured immediately upstream of the thermal sub-shock. }
\section{Summary and Discussion}
\begin{figure}[tb]
\centering
\includegraphics[width=0.99\linewidth]{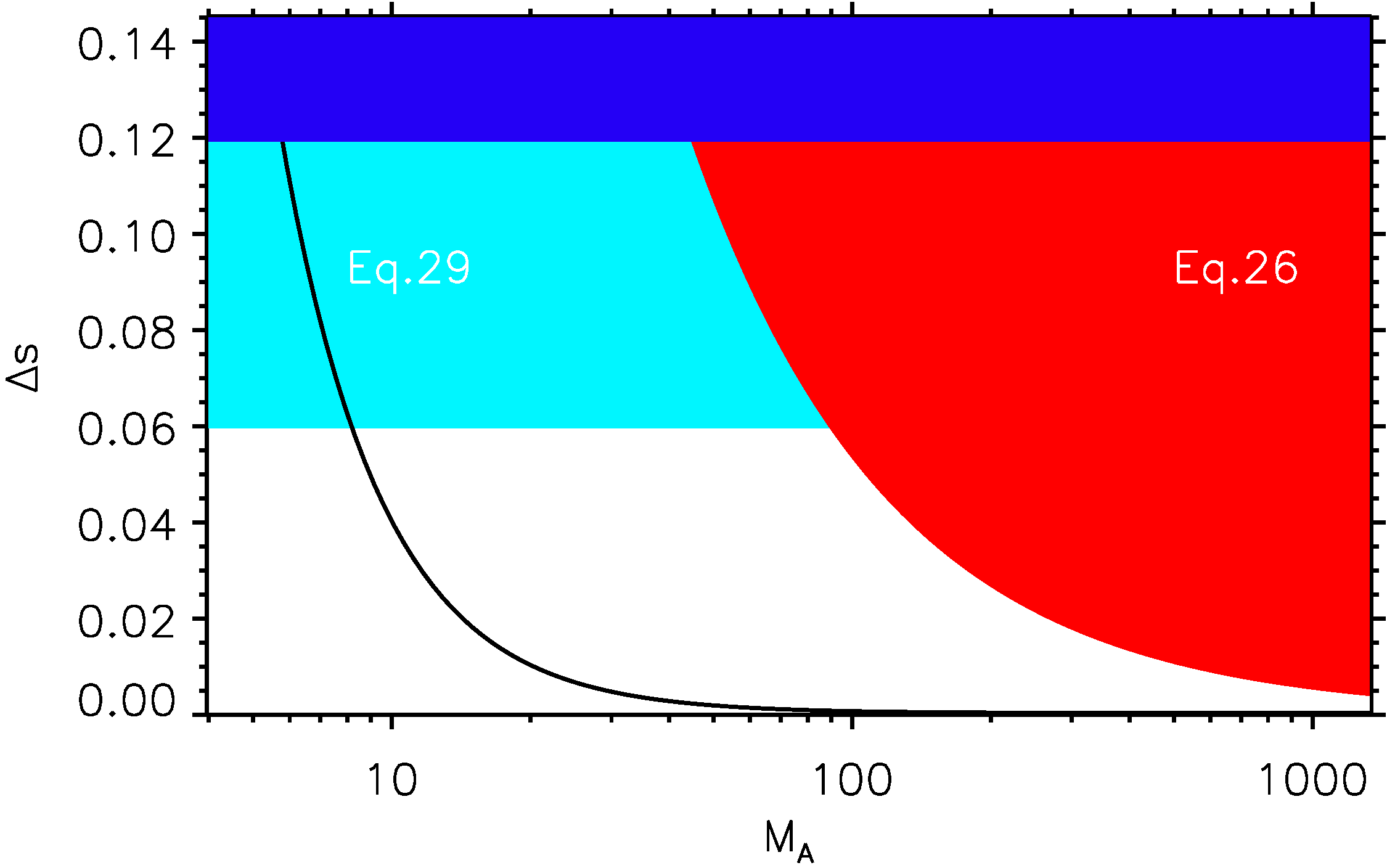}
\caption{Exclusion limits for the spectral softening, $\Delta s$, as \mpf{a} function of the Alfv\'enic Mach number, $\ma$, assuming $\eta=4$ and $\epsilon=0.5$. The red area \mpf{indicates a} violation of eq.~\ref{eq11a}. The cyan and blue areas are excluded by eq.~\ref{eq13} for $\vsh\ucr/c\ubu$ set to $10^{-3}$ and $4\cdot 10^{-3}$, respectively. The black solid line indicates the softening that is imposed by the magnetic pressure.}
\label{delta-s}
\end{figure}

We calculated the energy-transfer rate from cosmic rays to non-resonant plasma waves in the precursor of the forward shock of an SNR. Two different ways of calculation led to essentially the same result for the softening of the particle spectra that is imposed by that energy transfer. \mpo{It can be related to the energy-density ratio of amplified turbulent magnetic field and bulk plasma flow.} \mpon{That ratio is at most the inverse square of the Alfv\'enic Mach number of the thermal sub-shock, which we write with the full turbulent field amplitude,
\begin{equation}
\ma=\sqrt{\frac{\ubu}{U_{\delta B}+U_{B_0}}}\mponn{=
\sqrt{\frac{\ubu}{U_{\delta B}}}\frac{1}{\sqrt{
1+\frac{U_{B_0}}{U_{\delta B}}}} }.
\label{eq11aa}
\end{equation}
For \mpf{a} turbulently amplified magnetic field, $U_{\delta B}\gg U_{B_0}$, the spectral steepening can hence be written either with the Alfv\'enic Mach number of the sub shock or with the energy density of the turbulent field immediately upstream of the shock.} The most important finding is then that for Bohm scaling of diffusion in the precursor with factor $\eta$, the change in spectral index is invariably \mpo{
\begin{equation}
\Delta s
\lesssim\frac{2\,(s-1)^2\,\eta}{3\,\epsilon\,\ma}=\mpon{\frac{2\,(s-1)^2\,\eta}{3\,\epsilon}\sqrt{\frac{U_{\delta B}}{\ubu}}} .
\label{eq11a}
\end{equation}
Here $s$ is the cosmic-ray spectral index without steepening, and $\epsilon$ is the magnetic fraction of the \mpf{turbulence} energy density immediately upstream of the shock.}  
In Figure~\ref{delta-s} we show this constraint as red exclusion area in a display of spectral softening vs. $\ma$.

The magnetic-field strength immediately upstream of the shock, and hence $\ma$, cannot be easily determined through observations. Essentially all radiation is produced downstream of the shock, and there are numerous processes operating at the shock that generate magnetic fields and can cause a jump in $\mathrm{rms}(B)$ that is much larger than the compression imposed at the shock \citep[e.g.][]{2007ApJ...663L..41G,2013ApJ...770...84F}. The true Alfv\'enic Mach number would then be larger than naively estimated.

The scale factor of the diffusion coefficient, $\eta$, could be large, but not too large at very high energies, otherwise a young SNR could not accelerate particles to that energy. Simulations suggest that $\eta\approx 4$, if only the nonresonant mode is driven \citep{2013MNRAS.430.2873R,2017MNRAS.469.4985K}. Waves driven by cosmic rays in one energy band also scatter cosmic rays with different \mpf{energies}, albeit with somewhat modified efficiency \citep{2008MNRAS.386..509R,2014ApJ...788..107B}, which suggests that $\eta\approx 4$ is also a reasonable estimate when one considers the entire spectrum of cosmic rays accelerated at the shock. \mpon{Modes driven by cosmic rays of the highest energies may provide large-angle scattering for lower-energy cosmic rays, for which the size of individual magnetic filaments is commensurate with the Larmor radius in those filaments.} Any other MHD mode, \mpon{including those that are nonlinearly driven by Bell's mode, could lower $\eta$, because it may provide cosmic-ray scattering and can reduce the Larmor radius, $r_\mathrm{L}$, through further magnetic-field amplification}. Likewise, MHD processes that generate large-scale magnetic field in the precursor \citep[e.g.][]{2009ApJ...707.1541B,2011MNRAS.410...39B,2016MNRAS.458.1645D,2017ApJ...850..126X} would reduce $\kappa$ and hence $\Delta s$. 

A remaining possibility is that the magnetic field in the precursor is not amplified, in which case one might expect $\eta \approx B^2/(\delta B)^2$. \mpo{Inserting \mpf{this into} eq.~\ref{eq11} and using eq.~\ref{eq11aa} again yields $\Delta s$ of the order of $\ma^{-1}$.} We conclude that $\eta$ can likely not significantly increase the level of spectral softening. 

\citet{2019MNRAS.488.2466B} based \mponn{their} estimate on the energy flux of turbulence and cosmic rays through the shock, and the finding $\Delta s\propto \vsh$ is a consequence of the assumed saturation level $U_{\delta B}\propto \vsh$. Hadronic gamma-ray emission from SNRs shows soft spectra mainly for older SNRs whose shocks have decelerated to about $0.01 c$ or less, for which eq.~\ref{eq:bell1} yields negligible steepening. Radio data are much more abundant, but they reflect electron spectra. Technically, ultrarelativistic electrons can also drive Bell's mode, albeit with opposite circular polarization \citep{2009ApJ...699..990B}. \mpo{As cosmic-ray ions typically carry more energy density than do the electrons, they would provide most of the amplified field, and it is questionable whether any estimate of spectral steepening can be directly applied to electrons and hence to radio synchrotron spectra. Besides, the youngest known SNR in the Galaxy, G1.9+0.3, has a very fast forward shock \citep{2008ApJ...680L..41R}, but the radio emission most likely comes from the reverse shock whose speed is considerably lower \citep{2019A&A...627A.166B}. } 

\mpo{\mponn{It is instructive to assess the energy losses of cosmic-ray electrons.} The ratio of the energy-loss time of protons (Eq.~\ref{eq5}) and the synchrotron loss time of electrons at the same energy is independent of the magnetic-field strength. The nonresonant streaming instability operates only upstream of the shock, whereas the energy loss electrons incur in the compressed magnetic field during the downstream half-cycle of shock acceleration is about ten times that of the upstream half-cycle. Then, for parameters typical for young SNRs, a shock speed $\vsh=7000\,\mathrm{km\,s^{-1}}$, $\opp= 1\,\mathrm{kHz}$, and $\np=0.5\,\mathrm{cm^{-3}}$, the energy-loss time ratio exceeds unity for 15-TeV particles and increases $\propto \gcr^2$. \mponn{At 50 TeV the synchrotron loss timescale of electrons would be 10\% of the proton loss time for driving Bell's mode. Electron acceleration to 50 TeV or more would \mpf{imply} that the acceleration timescale \mpf{is} still significantly shorter than the synchrotron loss time, and so the energy loss of cosmic-ray protons is negligible at this energy}. In other words,} if intense nonthermal multi-keV emission is produced, the energy loss, and hence the spectral steepening, is very small for hadronic cosmic rays that produce TeV-band gamma-ray emission.

Instead of a fixed saturation level, we consider the energy transfer throughout the cosmic-ray precursor, which accounts for the limited available time and includes energy that nonlinearly passes to heating or other types of turbulence, but requires the assumption that the linear growth rate of the non-resonant mode is also valid in the nonlinear stage. \mpo{A consistency check is provided by mandating that in the steady state the wave-driving power per unit shock area (eq.~\ref{eq9}) must be at least as large as the escape flux through the shock, $\vsh U_{\delta B}/\epsilon$. This yields an implicit lower limit on the cosmic-ray density that with eq.~\ref{nexp} corresponds to requiring at least one half exponential growth cycle for the nonresonant mode, 
\begin{equation}
\ucr \ge \frac{6}{\eta}\,\frac{\ubu}{\ma} \quad\Rightarrow\ N_\mathrm{exp}\gtrsim 0.5 .
\label{eq11b}
\end{equation}
Evidently that condition is met for any significant growth of the mode. As the mode has to grow from small fluctuations to a level $\delta B \gg B_0$ , one would instead need around ten exponential growth cycles. Interestingly, we recover the result of \citet{2019MNRAS.488.2466B}, $\Delta s\vert_\mathrm{Bell}\approx 4\,U_{\delta B}/(\epsilon U_\mathrm{cr})$, if we assume a cosmic-ray density that equals the right-hand side eq.~\ref{eq11b}. In that case $N_\mathrm{exp}\simeq 0.5$, meaning there is no time for the mode to grow, and so the steady-state level $U_{\delta B}/U_\mathrm{cr} \approx \vsh /(2c)$ cannot be reached.}

\mpo{Our calculation of the number of exponential growth times in eq.~\ref{nexp} implicitly assumes that the mean \mponn{diffusion coefficient,} $\kappa$, in the cosmic-ray precursor is not much different from that immediately upstream of the shock, at least for cosmic rays at energies well below the cut-off. We noticed in section~\ref{sec2.2} that this must by roughly true, otherwise most upstream cosmic rays would reside very far from the shock and effectively escape, which is a far more serious loss process than wave driving.}

\mpon{The spectral modification implied by Eq.~\ref{eq11a}} does not explicitely depend on the energy density \mpf{of} cosmic rays, $\ucr$. The spectral softening reflects the per-particle rates of energy gain and escape from the shock. The energy transfer rate from cosmic rays to turbulence scales linearly with the cosmic-ray density, but normalized to the individual cosmic ray it is independent. Hence the independence of $\Delta s$ on the cosmic-ray density.  
We calculated the level of spectral softening ignoring the condition $k r_\mathrm{L}\gg 1$, under which the non-resonant mode can be driven. For $\delta B \gg B_0$ this relation can be rephrased as 
\begin{equation}
\frac{\vsh}{4c}\frac{\ucr}{U_{\delta B}} \gg 1 .
\label{eq12}
\end{equation}
\mpon{This condition requires that the left-hand side be much larger than unity, but how much larger? Let us conservatively suppose that is it larger than or equal to two. Inserting that into eq.~\ref{eq11a} we find}
\begin{equation}
\mpon{\Delta s \lesssim \frac{(s-1) \eta}{3\sqrt{2}\, \epsilon} \sqrt{\frac{\vsh}{c}\frac{\ucr}{\ubu}} }.
\label{eq13}
\end{equation}
This condition is also displayed in Figure~\ref{delta-s} for two rather high values of $\vsh$ and $\ucr$. \mpon{It appears as a constant limit at all $M_A$, although some values of $M_A$ may not be reached, at least not by magnetic-field amplification through Bell's mode.}

Given that the standard jump condition increases $\sqrt{U_{\delta B}/\ubu}$ in the downstream region to about 8 times its value in the upstream region, a strong magnetic field, \mpo{or small $\ma$,} will by itself modify the shock \citep[e.g.][]{1999A&A...343..303V,2000JPlPh..64..459L,2009MNRAS.395..895C} and hence soften the particle spectrum in a similar way as does the energy loss by driving turbulence. The solid black line in Figure~\ref{delta-s} indicates the magnitude of this effect \mpo{for a quasi-perpendicular shock with dynamically relevant field amplitude of $\sqrt{2/3}\,\delta B$. There is a marginal dependence of the curve on the shock-parallel magnetic-field component that we ignore here. Note also that \mpf{a} turbulently oriented magnetic field will for \mponn{$\ma\lesssim 20$} induce vorticity at the shock that can drive a turbulent dynamo in the downstream region and further amplify the field there.} 

\mpon{In conclusion, we find and show in Figure~\ref{delta-s}} that even for very efficient cosmic-ray acceleration, for which $\eta\approx 4$, and \mpon{the highest magnetic-field amplification that is allowed for Bell's mode}, the spectral softening appears to be \mpon{moderate}, $\Delta s \lesssim 0.1$, and it is negligible for standard SNR parameters. \mpon{As we explicitely allow for spatial variation in the cosmic-ray precursor, the shock speed in Eq.~\ref{eq13}, $\vsh$, is that of the thermal subshock, not that measured in the far-upstream frame.} \mpo{This statement is based on the energy transfer that can be accomplished within the time available, and it does not assume that a certain saturation level of the wave energy density is reached.}

\bibliography{eloss}{}
\bibliographystyle{aasjournal}

\end{document}